

\input harvmac

\def\VEV#1{\left\langle #1\right\rangle}
\def\sec{\ifmmode \,\, {\rm sec} \else sec \fi}
\def\eV {\ifmmode \,\, {\rm eV} \else eV \fi}
\def\keV{\ifmmode \,\, {\rm keV} \else keV \fi}
\def\MeV{\ifmmode \,\, {\rm MeV} \else MeV \fi}
\def\GeV{\ifmmode \,\, {\rm GeV} \else GeV \fi}
\def\TeV{\ifmmode \,\, {\rm TeV} \else TeV \fi}
\def\pbarn{\ifmmode \,\, {\rm pb} \else pb \fi}
\def\deg{\ifmmode ^\circ\, \else $^\circ\,$ \fi}
\def\Mx{{m_{\tilde\chi}}}

\def\Msq{m_{\tilde q}}

\def\fun#1#2{\lower3.6pt\vbox{\baselineskip0pt\lineskip.9pt
  \ialign{$\mathsurround=0pt#1\hfil##\hfil$\crcr#2\crcr\sim\crcr}}}

\def\neut{{\tilde\chi}}


\overfullrule=0pt


%
%
\xdef\prenum{\vbox{\hbox{IASSNS-HEP-94/14}\hbox{CWRU-P3-94}
\hbox{UCRL-JC-116940}
\hbox{February 1994}
\hbox{revised May 1994}}}

\Title{\prenum}{ \vbox{ \centerline{\hbox{Implications of Recent
Nucleon Spin Structure Measurements}}
\centerline{\hbox{ For Neutralino Dark Matter Detection}} }}
\def\lmkadd{\it $^2$Departments of Physics and Astronomy, Case
Western Reserve
University, Cleveland, OH 44107-7079}
\def\mtradd{ \it $^3$P-Division/Physical Sciences Directorate,
Lawrence Livermore National Laboratory, Livermore, CA 94550}
\def\mtraddtwo{ \it $^4$Institute of Geophysics and Planetary
Physics, Lawrence Livermore National Laboratory, Livermore, CA 94550}
\def\mkadd{\it $^1$School of Natural Sciences, Institute for Advanced
Study,
        Princeton, NJ 08540}

\centerline{Marc
Kamionkowski$^1$\footnote{$^\dagger$}{kamion@guinness.ias.edu},
Lawrence M.
Krauss$^2$\footnote{$^\ddagger$}{krauss@genesis1.phys.cwru.edu}, and
M.
Ted Ressell$^{3,4}$\footnote{$^*$}{ressell@sunlight.llnl.gov}}
\medskip
\centerline\mkadd
\centerline\lmkadd
\centerline\mtradd
\centerline\mtraddtwo

\vskip 0.5in
\centerline{ABSTRACT}

Predicted rates for direct and indirect detection
of dark-matter neutralinos depend in general on the spin content
of the nucleon.  Neutralinos that are predominantly $B$-ino are
the likeliest candidates for detection via spin-dependent
interactions.  Uncertainties in the measured spin content of the
nucleon may lead to dramatic uncertainties in the rates for
detection of $B$-inos by scattering off of nuclei with unpaired
neutrons.  Rates for spin-dependent scattering of $B$-inos off
of nuclei with unpaired protons are far more robust, as are
rates for capture of $B$-inos in the Sun.

\vfill
\Date{}


%
\lref\darkmatter{See, for example,V.~Trimble, \sl Ann.
        Rev. Astron. Astrophys. \bf 25\rm, 425 (1989); J.~R.~Primack,
       B.~Sadoulet, and D.~Seckel, \sl Ann. Rev. Nucl. Part. Sci.
       \bf B38\rm, 751 (1988);  \sl Dark Matter in the Universe,
       \rm eds.  J.~Kormendy and G.~Knapp (Reidel, Dordrecht,
       1989).}
\lref\slac{D. L. Anthony et al, {\sl Phys. Rev. Lett.} {\bf 71}, 959
(1993)}
\lref\haberkane{H.~E.~Haber and G.~L.~Kane,
        \sl Phys. Rep. \bf 117\rm, 75 (1985).}
\lref\ellishag{J.~Ellis, J.~S.~Hagelin, D.~V.~Nanopoulos,
K.~A.~Olive,
       and M.~Srednicki, \sl Nucl. Phys. \bf B238\rm, 453
       (1984); K.~Griest, \sl Phys. Rev. \bf D38\rm, 2357 (1988);
        FERMILAB-Pub-89/139-A (E); \sl Phys. Rev.
        Lett. \bf 61\rm, 666 (1988).}
\lref\heavy{K.~Griest, M.~Kamionkowski, and M.~S.~Turner, \sl
Phys. Rev. D \bf 41\rm, 3565 (1990); K.~A.~Olive and M.~Srednicki,
        \sl Phys. Lett. \bf B230\rm, 78 (1989);
        K.~A.~Olive and M.~Srednicki, \sl Nucl. Phys. \bf
        B355\rm, 208 (1991).}
\lref\witten{M. W. Goodman and E. Witten, {\sl Phys. Rev. D}
{\bf 31}, 3059 (1985); I. Wasserman, {\sl Phys. Rev. D} {\bf
33}, 2071 (1986).}
\lref\neutrinos{For a review, see M. Kamionkowski in {\it High Energy
Neutrino Astrophysics}, edited by V.~J.~Stenger,
       J.~G.~Learned, S.~Pakvasa, and X.~Tata (World Scientific,
       Singapore, 1992), p. 157}
\lref\experiments{For reviews of current and future
       energetic-neutrino detectors, see, e.g., {\it High Energy
       Neutrino Astrophysics,} edited by V.~J.~Stenger,
       J.~G.~Learned, S.~Pakvasa, and X.~Tata (World Scientific,
       Singapore, 1992).}
\lref\emc{A. Ashman et al, {\sl Phys. Lett. B.} {\bf 206}, 364
(1988);
R. L. Jaffe and A. Manohar, {\sl Nucl. Phys.} {\bf
B337}, 509 (1990).}
\lref\smc{The Spin Muon Collaboration: D. Adams et al.,
CERN-PPE/94-57.}
\lref\nqm{L. M. Krauss, P. Romanelli, {\sl Phys. Rev. D} {\bf 39},
1225 (1989)
R. Flores, K. A. Olive, and M. Srednicki, {\sl Phys.
Lett. B} {\bf 237}, 72 (1990).}
\lref\leszek{L. Roszkowski, {\sl Phys. Lett. B} {\bf 262},
59 (1991).}
\lref\ressell{M. T. Ressell et al., {\sl Phys. Rev. D} {\bf 48},
5519 (1993).}
\lref\lmk{F. Iachello, L. M. Krauss, and G. Maino,
{\sl Phys. Lett. B.} {\bf 254}, 220 (1991)}
\lref\nuclear{For a review, see J. Engel, S. Pittel, and P.
Vogel, {\sl Int. J. Mod. Phys. E} {\bf 1}, 1 (1992).}
\lref\DRS{M. Drees, and M. M. Nojiri, {\sl Phys. Rev. D} {\bf 48},
3483 (1993)}
%

%
%

There is almost universal agreement among astrophysicists on
the existence of dark matter in our galactic halo \darkmatter.  The
nature of this nonluminous matter is perhaps the most important
unsolved
problem in cosmology and particle physics.  One of the leading
candidates for the dark matter is the neutralino, a linear
combination of the supersymmetric partners of the photon, $Z^0$
boson, (or the $U(1)$ gauge field $B$ and the third component of
the $SU(2)$ gauge field $W_3$) and two neutral Higgs bosons
\haberkane\ellishag\heavy:
\eqn\neutralino{
\neut=Z_1 \tilde B + Z_2 \tilde W_3 + Z_3 \tilde H_1 + Z_4
\tilde H_2,}
where the tildes denote superpartners, and the $Z_i$ are
coefficients which determine the composition of the neutralino.

A variety of complementary
experimental avenues are being pursued in an effort to discover
neutralinos in our halo.  The first of two of the most promising
techniques involves direct detection of the recoil energy
imparted to a nucleus in a low-background detector from elastic
scattering of a halo neutralino off the nucleus \witten.  The second
involves indirect detection via observation of energetic
neutrinos from annihilation of neutralinos that have been
captured in the Sun \neutrinos\experiments.

Generally, neutralinos can scatter off of nuclei either through
an axial-vector (or spin-dependent) interaction, a scalar (or
spin-{\it in}dependent) interaction, or both.  Here we discuss
only the axial-vector interaction.  The matrix element for
spin-dependent neutralino-nucleus elastic scattering depends on
the nuclear
matrix elements of the axial-vector $u$-, $d$-, and $s$-quark
operators.  These are easily calculated using the $SU(3)$ naive quark
model (NQM) \witten \nqm.  However EMC measurements of the spin
structure function of the proton suggested
that the quark contribution to the spin of the proton was suppressed,
so that the NQM estimates would have to be altered
\emc.  As a result, most of the recent calculations of
neutralino-detection rates assume EMC values
for the spin content of the nucleon.

Recent new measurements of the proton
spin structure function by the Spin Muon Collaboration (SMC)
find values close to the EMC results \smc, but data from the E142
experiment \slac\ at SLAC agrees very well with the NQM predictions.
(The latter data is in disagreement with the fundamental
Bjorken sum rule, however.)  Our purpose in
this paper is not to advocate that either of these new data
is a definitive result; rather, we wish to point out the
uncertainties they introduce
for direct- and indirect-detection rates in existing and
proposed detectors. We demonstrate that
they imply that the predicted event rates could be altered by
more than an order of magnitude in certain cases.

The effects of the spin structure of the nucleon on detection
rates are well known.  Here we first argue that the
likeliest candidate for detection via
spin-dependent scattering is a $B$-ino
($Z_1=1$ and $Z_i=0$ for $i\neq1$). In this case the
implications of existing uncertainties can be explicitly explored.
There is a potential
cancelation in the contributions of the various quarks to the
matrix element for scattering of a $B$-ino from a neutron, so
small uncertainties in the measured spin content of the nucleon
can lead to dramatic variations in the predicted rates for
elastic scattering of $B$-inos off of nuclei with unpaired
neutrons.  There is no such cancelation for scattering from
protons, so rates for detection in detectors with
unpaired-proton nuclei are much more robust.

If the neutralino
is primarily Higgsino ($Z_1,Z_2\ll Z_3,Z_4$), spin-dependent
interactions are generally suppressed, so to a large extent,
Higgsinos will be invisible to the detection techniques
discussed here.  If the neutralino is a
mixed gaugino/Higgsino state, the neutralino-nucleus interaction
is primarily a scalar interaction; the spin-dependent
interaction is generally subdominant.  If the neutralino is
predominantly gaugino, then it is most often a $B$-ino,
and $B$-inos interact with nuclei only through a spin-dependent
interaction (unless the squark mass is close to the neutralino
mass or there is significant squark mixing \DRS).
Therefore, $B$-inos are the most likely candidates
for detection via experiments that depend on the spin-dependent
neutralino-nucleus interaction.
Also note that as accelerator experiments raise the
lower bounds to the neutralino mass, the fraction of parameter
space where the neutralino is either a pure Higgsino or pure
$B$-ino increases \heavy.  In addition, there are
theoretical indications that the neutralino is most likely a
$B$-ino \leszek.

Complete expressions for the spin-dependent neutralino-nucleus
elastic-scattering cross section, $\sigma_{\rm SD}$, for an
arbitrary neutralino can be found, for example, in
Ref.~\ressell.  Here, we introduce those parts of the
calculation that are relevant to our central arguments.
The spin-structure dependence of the cross section may be written
\eqn\sigmaSD{
\sigma_{\rm SD} \propto \Lambda^2 J(J+1),}
where $J$ is the total nuclear angular momentum.  The matrix
element, $\Lambda$, depends in particular on the neutralino
interactions and on nuclear matrix elements of the quark-spin
operators.

Although quite sophisticated calculations of the
nuclear matrix elements have been performed \ressell\lmk\nuclear,
for the purposes of
illustration, we will first use the simplest, the
independent single-particle shell model (ISPSM) of the nucleus. This
will give a fairly accurate indication of the magnitude of the
effects investigated here.  In this model, the
nuclear spin is contributed entirely by a single unpaired
proton (neutron), and $\Lambda=a_{p(n)} s_{p(n)}/J$, where
$s_{p(n)}=1/2$ is the spin of the unpaired proton (neutron).  (To
clarify
a common misconception, note that
$J$ appears in the denominator of $\Lambda$.  Therefore, the
cross section is {\it not} proportional to the square of the
total nuclear angular momentum.)  It can be shown that for a
pure $B$-ino (and in the limit of large squark masses $\Msq\gg \Mx$)
\ressell,
\eqn\bino{
a_{p} \simeq 0.303 {m_W^2 \over \Msq^2} \left[ {17 \over 36}
\Delta u  + {5\over 36} (\Delta d + \Delta s) \right],}
where $m_W=80$~GeV is the $W^\pm$-boson mass, and the $\Delta
q$'s are the proton (neutron) quark-spin matrix element.
The coupling $a_n$ is obtained by interchanging $\Delta u$ and
$\Delta d$ in the above expression.

In terms of the quantities determined by measurements of the
nucleon spin structure functions (that is, the first moment of
the proton spin-dependent structure function, $\Gamma$, the
total quark spin contribution to the nucleon spin, $\Delta
\Sigma$, and $\Delta s$), the $u$ and $d$-quark matrix elements
are $\Delta u = 6\Gamma - (1/3)\Delta \Sigma$ and $\Delta d = (4/3)
\Delta \Sigma - 6\Gamma - \Delta s$ (neglecting small
corrections due to the running of the strong coupling constant).
Thus, the $B$-ino
couplings to protons and neutrons are
\eqn\apcoupling{
a_p \propto 72 \Gamma + \Delta \Sigma,}
and
\eqn\ancoupling{
a_n \propto 21 \Delta \Sigma - 72 \Gamma - 12 \Delta s.}
The coupling of $B$-inos to protons depends almost entirely
on $\Gamma$, which is fairly well determined.  On the other
hand, $a_n$ depends on a difference of three measured
quantities.  There is potentially a cancelation in the three
terms that contribute to $a_n$ which can lead to a rather large
uncertainty in $a_n$ even if the spin content of the nucleon is
fairly well determined.

{\nobreak
{
$$
\vbox{
\centerline{Spin Matrix Elements}
\medskip
\halign{# \hfill & \quad # \hfill & \quad # \hfill & \quad # \hfill &
\quad #
\hfill & \quad # \hfill & \quad # \cr
\noalign{\hrule}
\noalign{\smallskip}
\noalign{\hrule}
\noalign{\medskip}
 & NQM  & EMC & SMC & All & All$_{\Delta \Sigma +
 1 \sigma}$ \cr
\noalign{\medskip}
\noalign{\hrule}
\noalign{\medskip}
$\Gamma$ & 0.188 & 0.137 & 0.136 & 0.145 & 0.145\cr
$\Delta \Sigma$ & 0.60 & 0.12 & 0.22 & 0.30 & 0.42\cr
$\Delta s$ & 0 & -0.16 & -0.12 & -0.09 & -0.09 \cr
\noalign{\medskip}
\noalign{\hrule}
\noalign{\medskip}
$a_p$ & 0.119 & 0.083 & 0.084 & 0.090 & 0.091\cr
$a_n$ & -0.0081 & -0.045 & -0.031 & -0.026 & -0.0045\cr
\noalign{\medskip}
\noalign{\hrule}
\noalign{\medskip}
$\Delta u$ & 0.93 & 0.78 & 0.74 & 0.77 & 0.73 \cr
$\Delta d$ & -0.33 & -0.50 & -0.40 & -0.38 & -0.22 \cr
\noalign{\medskip}
\noalign{\hrule}
}}
$$
\noindent{Table 1.  Quark spin content of the proton, and $B$-ino
spin-dependent nucleon coupling constants, determined from the
$SU(3)$ naive quark model (NQM) \witten\nqm\ and for measured
spin-dependent structure functions
from EMC \emc, SMC \smc, and a compilation (All) \smc.  Also
listed are values using the $1\sigma$
error on $\Delta \Sigma$ from the compilation.   A factor of
$m_W^2/\Msq^2$ is implied in the values of $a_n$ and $a_p$.
\bigskip
}

In Table 1, we list, for purposes of comparison, several reported
values for $\Gamma$, $\Delta \Sigma$, and $\Delta s$, as well as
the resulting values for $\Delta u$ and $\Delta d$, and for $a_p$
and $a_n$ (scaled by
$m_W^2/\Msq^2$).  We list results from the $SU(3)$ naive quark
model (NQM) \witten\nqm, the old EMC results \emc, new results
from SMC (\smc), and a common
evaluation from all existing data (All) \smc.  We also allow for a
reported uncertainty of $0.12$ ($1\sigma$ statistical plus
systematic added in quadrature) in the value for $\Delta \Sigma$
from the
compilation and list the resulting values in the last column of
Table 1.  The E142 results remain controversial since they do
not satisfy the Bjorken sum rule, we do not include them in the
Table, but we point out that they seem to be closer to the NQM
than the EMC results.

The interaction rate is proportional to $a_{p(n)}^2$, and as the
Table illustrates, the different determinations of $a_n$ vary
widely, while the determinations of $a_p$ are far more robust.
For example, for scattering off of nuclei with an unpaired
neutron (e.g.
$^{73}$Ge or $^{29}$Si), the rate based on the compiled data could be
decreased by roughly 1/2 to 1/100 compared to the rate
based on EMC.  For scattering off of nuclei with an unpaired proton
(e.g. $^{39}$K, $^{93}$Nb, as well as hydrogen), the
compiled-data rate
would be slightly larger than the EMC rate. On the other hand,
the rate for scattering off unpaired-neutron nuclei based on the NQM
is roughly $ 1/30 $ the rate obtained
using EMC values, while the rate for scattering off of
unpaired-proton nuclei is roughly twice as large.
Although these results do not necessarily apply to the most general
neutralino, they do imply that the
sensitivities of detectors with unpaired-neutron nuclei to
dark-matter neutralinos with spin-dependent interactions can vary by
more than order of magnitude.  In fact, if we take the
compiled-data values for $\Gamma$ and $\Delta s$ and take
$\Delta \Sigma = 0.45$ (slightly greater than 1 $\sigma$ above the
common average),
then $a_n=0$; the cancelation is exact.  In other words,
the neutron--$B$-ino coupling could be consistent with zero
within the reported spin-structure uncertainties.
This example serves to
illustrate the potentially large uncertainties in the
$B$-ino--neutron coupling that arise from rather well determined
spin structure functions.  On the other
hand, at least for $B$-inos, the sensitivity of detectors with
unpaired-proton nuclei does not vary greatly.

For $^{73}$Ge and $^{29}$Si, two target
nuclei that are used in existing and planned detectors,
 we re-do the calculation using
state-of-the-art estimates of the nuclear matrix element.  In this
way we also probe the sensitivity of our conclusions to
nuclear-physics uncertainties.  We follow the analysis of
Ref.~\ressell.  In a more general model,
\eqn\Lamdaeqn{
\Lambda= [a_p\VEV{S_p} + a_n \VEV{S_n}]/J,}
where the proton and neutron spin averages, $\VEV{S_p}$ and
$\VEV{S_n}$, are computed given a nuclear model.  For $^{29}$Si,
the best estimates are (from the shell model) $\VEV{S_p}=
-0.002$ and $\VEV{S_n}=0.13$.  Using these values, we find that
the neutralino-nucleus interaction rates obtained using the
central value from the compiled data could be about 1/3 of those
obtained using the EMC values, a fraction comparable to that
obtained with the single-particle shell model.

For $^{73}$Ge we again follow Ressell et al. \ressell\ and use a
quenched shell model.  To do so, we take $\VEV{S_n}=0.372$ and
$\VEV{ S_p}=0.009$. In this case the compiled-data rates
for neutralino interactions
with $^{73}$Ge would be only 1/3 of the EMC rates---again a
fraction comparable to that obtained with the ISPSM.

A very promising alternative to direct-detection experiments are
searches for energetic neutrinos from neutralino annihilation in the
Sun \neutrinos.  Particles such as $B$-inos which have only
spin-dependent
interactions (and no scalar interactions) with nuclei can be
captured in the Sun by scattering from hydrogen and
annihilate therein producing energetic neutrinos which may be
detected in astrophysical neutrino detectors.  The bulk of the
Earth's mass is in spinless nuclei, so such particles would {\it
not} be captured in the Earth.  The flux of neutrinos from the
Sun is proportional to the cross section for elastic scattering
off of protons, and our results suggest the predicted capture rates
will be fairly insensitive to variations at the level of the existing
experimental uncertainties.   This indirect-detection scheme will
provide a valuable alternative to searches with unpaired-neutron
detectors if the spin-structure uncertainties remain unresolved.

To conclude, we have given general arguments that the $B$-ino is
the likeliest candidate for detection involving a spin-dependent
interaction with nuclei.  If the new SMC values for the quark-spin
content of the nucleon are upheld, then the rates for
detection of $B$-inos in detectors with unpaired-neutron nuclei
could be suppressed relative to the rates obtained assuming EMC
values by potentially large factors.  On the other hand,
rates for scattering off of unpaired-proton nuclei, as well as
rates for energetic neutrinos from neutralino annihilation in
the Sun will not vary dramatically from previous estimates.
We have also shown that these conclusions are
insensitive to the specific nuclear model assumed.  Given
the tentative status of the current experimental
situation, the range in detection rates we have estimated here
should be considered as representative of how large the existing
uncertainties are, and also how small the detection
rates could realistically be.

We also note that predictions of rates in detectors with
unpaired-neutron nuclei depend much more sensitively on the
specific quark-spin matrix elements than do the rates in
detectors with unpaired-proton nuclei or the rates for energetic
neutrinos from the Sun.  So, it is possible that even if the
spin-structure measurements are made more precise, unpaired-proton
detectors and energetic-neutrino searches will still provide
less ambiguous information than unpaired-neutron detectors on
properties of dark-matter neutralinos.

As our work demonstrates, by far the largest uncertainty in
spin-dependent neutralino detection rates can come from the
uncertainty
in the spin structure of the nucleon (overwhelming nuclear-physics
and the halo-density uncertainties).  It will thus
be important to resolve the spin structure of the nucleon
before precise estimates of general neutralino detection rates will
be possible.

\bigskip

We thank F.~Wilczek for discussions, and A. Manohar for informing us
of the details of the SMC preliminary data.
M.K. was supported by the U.S. Department of
Energy under contract DEFG02-90-ER 40542.
Work at LLNL was performed under the auspices of the U.S.
Department of Energy under contract W-7405-ENG-48
and DOE Nuclear Theory Grant SF-ENG-48.

\listrefs
\bye